\documentclass[journal]{IEEEtran}
\usepackage{cite}
\usepackage{amsmath,amsfonts}
\usepackage{algorithmic}
\usepackage{algorithm}
\usepackage{array}
\usepackage[font=footnotesize]{subcaption}
\usepackage{textcomp}
\usepackage{stfloats}
\usepackage{verbatim}
\usepackage{graphicx}
\usepackage{balance}
\usepackage{subcaption}
\usepackage[dvipsnames]{xcolor}
\usepackage{cite}
\usepackage{tabularx}
\usepackage{booktabs}
\usepackage{acronym}

\usepackage{color}

\usepackage{tabu,longtable}
\usepackage{diagbox}
\usepackage{threeparttable}
\usepackage{makecell}
\usepackage[table]{xcolor}
\usepackage{multirow} 
\usepackage{tikz} 
\usepackage[utf8]{inputenc}
\usepackage{pgfplots}
\pgfplotsset{compat=1.18} 
\usepackage{units} 		
\usepackage{bm} 		
\usepackage{amssymb} 	

\acrodef{4G}{fourth generation}
\acrodef{3GPP}{Third Generation Partnership Project}
\acrodef{5G}{fifth generation}
\acrodef{5G-A}{5G-advanced}
\acrodef{6G}{sixth generation}
\acrodef{AGV}{automated guided vehicle}
\acrodef{A-IoT}{ambient IoT}
\acrodef{AmBC}{ambient backscatter communication}
\acrodef{AoA}{angle-of-arrival}
\acrodef{API}{application programming interface}
\acrodef{BD}{backscattering device}
\acrodef{BS}{base station} 
\acrodef{CN}{core network}
\acrodef{CRLH}{composite right/left-handed}
\acrodef{CRS}{Cell-Specific Reference Signal}
\acrodef{CW}{continuous-wave}
\acrodef{D2R}{device-to-reader}
\acrodef{dAPP}{distributed applications}
\acrodef{DL}{downlink}
\acrodef{DMRS}{Demodulation Reference Signal}
\acrodef{DSB}{double-sideband}
\acrodef{DU}{distributed unit}
\acrodef{EC}{European Commission}
\acrodef{EM}{electromagnetic}
\acrodef{eNB}{evolved nodeB (4G base station)}
\acrodef{FMCW}{frequency-modulated continuous wave}
\acrodef{gNB}{gNodeB (5G base station)}
\acrodef{HR}{human resources}
\acrodef{IC}{integrated circuit}
\acrodef{IoT}{Internet of Things} 
\acrodef{ISAC}{integrated sensing and communication}
\acrodef{ITU}{International Telecommunication Union}
\acrodef{LCA}{life cycle assessment}
\acrodef{LDO}{low-dropout}
\acrodef{LMF}{location management function}
\acrodef{MAC}{medium access control}
\acrodef{mMTC}{massive machine-type communications}
\acrodef{mmWave}{millimeter-wave}
\acrodef{MTC}{machine-type communication}
\acrodef{NAS}{non-access stratum}
\acrodef{nGRG}{next generation research group}
\acrodef{NOMA}{non-orthogonal multiple access}
\acrodef{OA}{open access}
\acrodef{OFDM}{Orthogonal Frequency Division Multiplexing}
\acrodef{OOK}{on–off keying}
\acrodef{O-RAN}{open-RAN}
\acrodef{OTA}{over-the-air}
\acrodef{PCB}{printed circuit board}
\acrodef{PoC}{proof of concept}
\acrodef{PV}{photovoltaic}
\acrodef{QT}{quantitative target}
\acrodef{R2D}{reader-to-device}
\acrodef{RAN}{radio access network}
\acrodef{rApp}{RANapplication}
\acrodef{RCS}{radar cross section}
\acrodef{RF}{radio frequency}
\acrodef{RH}{relative humidity}
\acrodef{RIC}{RAN intelligent controller}
\acrodef{R-SLAM}{radio-SLAM}
\acrodef{RU}{radio unit}
\acrodef{SAW}{surface acoustic wave}
\acrodef{SCEF}{Service Capability Exposure Function}
\acrodef{SDR}{software-defined radio}
\acrodef{SeCF}{sensing control function}
\acrodef{SIC}{successive interference cancellation}
\acrodef{SLAM}{simultaneous localization and mapping} 
\acrodef{SMO}{service management and orchestration}
\acrodef{SNR}{signal-to-noise ratio}
\acrodef{SoC}{system-on-chip}
\acrodef{SPF}{sensing processing function}
\acrodef{SPI}{serial peripheral interface}
\acrodef{SRS}{sounding reference signal}
\acrodef{TL}{transmission line}
\acrodef{TRL}{technology readiness level}
\acrodef{UC}{use case}
\acrodef{UC-ZED}{universal cellular zero-energy device}
\acrodef{UE}{user equipment}
\acrodef{UL}{uplink}
\acrodef{UWB}{ultra-wideband}
\acrodef{WG}{working group}
\acrodef{XG}{any generation (4G/5G/6G and beyond)}
\acrodef{XG-ZED}{generation-agnostic zero-energy device}
\acrodef{ZED}{zero-energy device}
\acrodef{AI}{artificial intelligence}
\acrodef{V2X}{vehicle-to-everything}
\acrodef{EIRP}{effective isotropic radiated power}
\acrodef{NF}{noise figure}
\acrodef{RFID}{radio-frequency identification}
\acrodef{BC}{backscatter communication}
\acrodef{FDSOI}{fully depleted silicon-on-insulator}
\acrodef{CRS}{common reference signal}
\acrodef{FSK}{frequency-shift keying}

\begin{document}

\title{Generation-Agnostic Zero-Energy Devices for Sustainable Connectivity, Sensing, and Localization}

\author{Navid~Amani, Filiberto~Bilotti, Davide~Dardari, Raffaele~D'Errico, Riku~Jäntti, Gianni~Pasolini, Dinh-Thuy~Phan-Huy, Davide~Ramaccia, Olivier~Rance, and~Henk~Wymeersch
\thanks{
N. Amani acknowledges the support from VINNOVA ZE-IoE project (Grant No. 2024-02416).
F. Bilotti acknowledges the support by the European Union under the Italian National Recovery and Resilience Plan (NRRP) of NextGenerationEU, partnership on “Telecommunications of the Future” (PE00000001 - program “RESTART”).
R.~J\"antti and H.~Wymeersch, R. D'Errico acknowledge support from the European Commission through the Horizon Europe/JU SNS projects AMBIENT-6G (Grant Agreement No.~01192113) and 6G-DISAC (Grant Agreement No.~101139130), and ANR French Project S²LAM.}
}

\maketitle

\begin{abstract} 
The massive scale of \ac{IoT} connectivity expected in 6G networks raises unprecedented challenges in energy use, battery waste, and lifecycle sustainability. Current cellular IoT solutions remain bound to the lifetime of underlying network generations and rely on billions of disposable batteries, creating unsustainable economic and environmental costs. This article proposes  \acp{XG-ZED}, a new class of backscatter-based IoT devices that are battery-less, spectrum-agnostic, and future-proof across successive network generations. XG-ZEDs exploit existing ambient wireless signals for communication, sensing, and localization, transforming infrastructure and user devices into universal enablers of ultra-low-power connectivity. We review architectural classifications, communication protocols, network integration, and representative applications such as sensing, localization, and radio-SLAM, while outlining the challenges ahead. 

\end{abstract}

\acresetall 
\vspace{-3mm}
\section{Introduction}
The evolution from current {4G} and {5G} networks to {6G} promises transformative advancements in wireless communication, particularly to meet the demands of massive-scale \ac{IoT} connectivity. As {6G} is expected to support device densities up to an order of magnitude higher than {5G}, energy efficiency emerges as a critical requirement for ensuring long-term sustainability. Existing \ac{MTC} standards, such as NB-\ac{IoT}, LTE-M, and RedCap, support low-cost, low-power devices with long battery life under favorable conditions. However, these technologies fall short for many emerging low-end \ac{IoT} applications that require ultra-low power, minimal cost, extended lifespans,  and operation in harsh environments without batteries  \cite{naser2023zero}. 

To overcome these limitations, \ac{A-IoT} based on \ac{BC} has emerged as a promising solution. By harvesting ambient energy, \ac{BC} enables battery-free, ultra-low-power connectivity, paving the way for sustainable and scalable \ac{IoT} deployments \cite{naser2023zero}. Moreover, \ac{BC} directly addresses a second major shortcoming of current cellular-based \ac{IoT}: their dependency on a single network generation (e.g., {4G} or {5G}). This dependency necessitates device replacement whenever spectrum is reallocated or operators migrate to a newer generation. \Ac{AmBC} enables a \textit{generation-agnostic approach to zero-energy \ac{IoT}}, ensuring devices are no longer constrained by specific cellular standards and can operate without onboard batteries, thus supporting truly unattended and long-term operation.
%
\ac{BC} has been the cornerstone of \ac{RFID}, 
where tags modulate information onto a continuous-wave signal generated by a dedicated reader. Despite the battery-free operation of passive \ac{RFID} tags, the 
communication range is typically restricted to less than 10 m, and the readers themselves are relatively costly and power-hungry, often requiring 0.5–2 W of transmit power~\cite{ aliasgari2020chipless}, preventing 
large-scale energy-efficient deployments. 

Recognizing this potential, the \ac{3GPP} has recently initiated study items focused on \ac{A-IoT}, addressing service requirements, potential use cases, and technical considerations within the \ac{RAN} \cite{butt2024ambient}. Early standardization efforts are now underway as part of \ac{3GPP} Release 19 with the TR38.769 on the study on solutions for \ac{A-IoT} in NR to enable large-scale deployment. In this context, ambient devices operate by harvesting energy from their surrounding environment, such as light, heat, motion, or RF signals, allowing them to function without traditional batteries.
While \ac{ZED} and AmBC have been studied extensively in terms of architectures, modulation schemes, and integration with specific standards, their operation and design from a generation-agnostic perspective have received limited attention. 



In this paper, we introduce the concept of \textit{\acp{XG-ZED}}, based on \ac{AmBC}, and analyze their architectural classifications from both hardware and software viewpoints. We further investigate network architectures and protocols for their integration. In addition, we explore emerging application scenarios, highlight the role of \acp{XG-ZED} in enabling scalable and ultra-low-power \ac{IoT} solutions, and conclude with a discussion of key challenges and future research opportunities in this rapidly evolving field.

\begin{figure*}[!t]
\centerline{\includegraphics[width=0.9\textwidth, trim=25cm 9cm 19cm 0.5cm, clip]{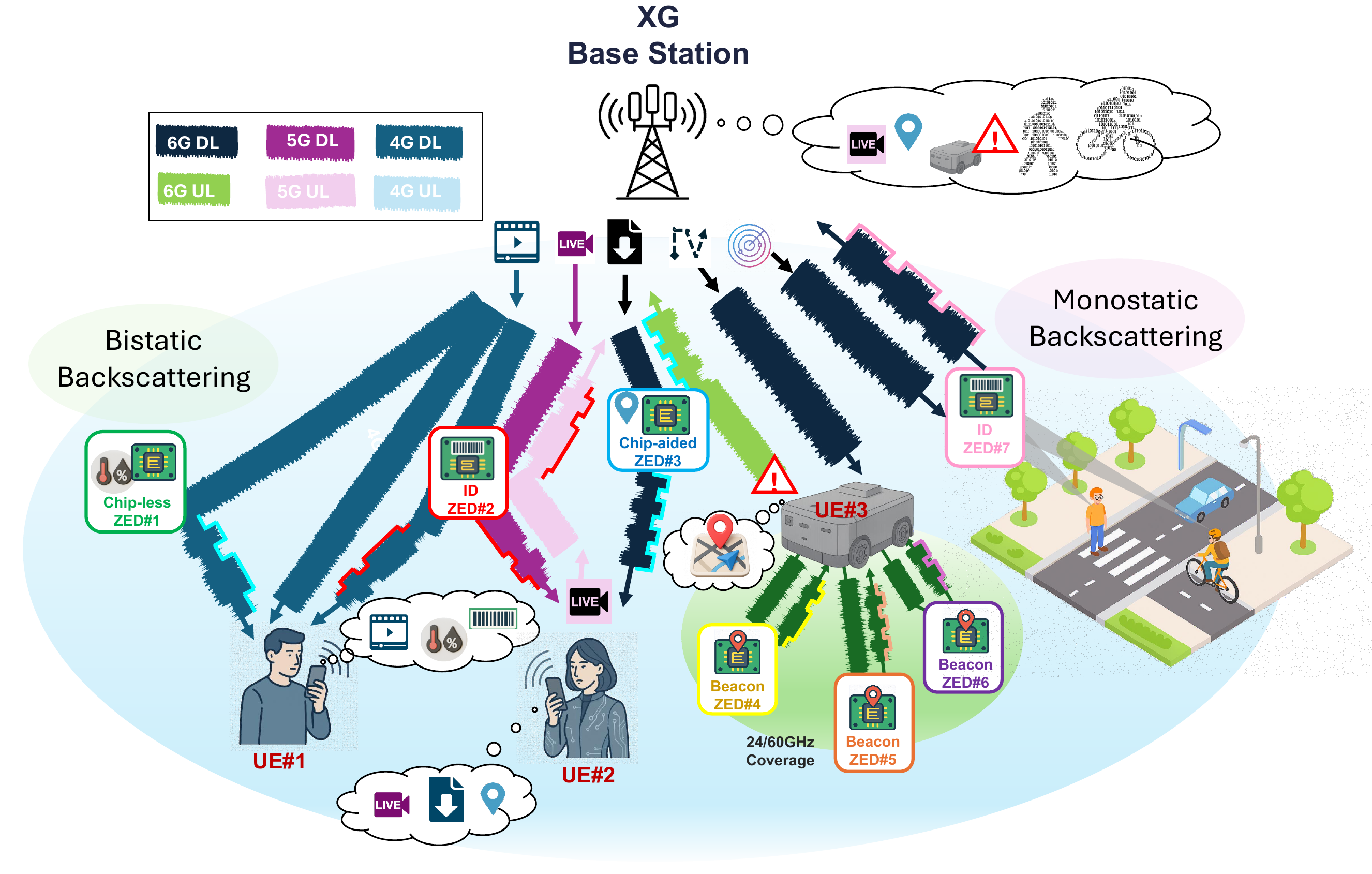}}
    \caption{Illustration of a multi-generation ambient backscatter system where \acp{XG-ZED} reuse 4G/5G/6G uplink and downlink signals. The XG base station supports monostatic reading, while user equipments enable bistatic operation. Different \ac{XG-ZED} types (chip-less, chip-aided, beacon) coexist across sub-6 GHz and mmWave bands, with retrodirective arrays extending coverage at 24/60 GHz.}
    \label{fig:Multi_Static} \vspace{-4mm}
\end{figure*}

\begin{table*}[t]
\caption{Key techniques and design aspects in ambient backscatter communication.}
\centering
\renewcommand{\arraystretch}{1.2}
\begin{tabularx}{\textwidth}{|l|X|X|X|}
\hline
\textbf{Parameter} & \textbf{Categories / Options} & \textbf{Role in \ac{A-IoT}} & \textbf{Examples / Notes} \\
\hline
\textbf{Backscatter Setup} & Monostatic, bistatic, multi-static & Defines how illuminator, \acp{ZED}, and interrogators are positioned & Bistatic used in ambient backscatter, \ac{CW} based schemes often monostatic \\
\hline
\textbf{Modulation Scheme} & OOK, ASK, FSK, PSK, and QAM. & Determines spectral efficiency, complexity, BER & ASK used in EPC Gen2; CSS in LoRa backscatter; \ac{OOK} in \ac{A-IoT} Rel 19. \\
\hline
\textbf{Ambient Waveform} & \ac{CW}, \ac{OFDM}, and FMCW & Affects the compatibility and decoding complexity & \ac{OFDM} in most ambient backscatter proposals\\ 
\hline
\textbf{Antenna Polarization} & Linear, circular, dual & Affects \ac{ZED}-interrogator coupling and range & Polarization diversity enhances robustness \\
\hline
\textbf{Interrogator Duplexing} & Half, Full & Affects timing and synchronization between transmission and reception & Full-duplex possible with separate channels \\
\hline
\textbf{Multiple Access} & ALOHA, FDMA, TDMA, CDMA, CSMA & Handles collisions among multiple \acp{ZED} & \ac{A-IoT} Rel. 19 uses FDMA, EPC uses slotted ALOHA\\ 
\hline
\textbf{Energy Source} & RF, solar, thermal, mechanical & Critical for zero-energy operation, RF energy harvesting limits range & Typically RF; Ambient backscatter must use other\\
\hline
\textbf{BD Type} & Passive, semi-passive, active, chip-aided, and chip-less & Defines power autonomy and complexity & Passive tags dominate in supply chain \\
\hline
\textbf{Coding Technique} & None, Hamming, Convolutional codes, LDPC, Polar & Enhances error correction and improve range & LDPC in modern ambient \ac{IoT} trials, Convolutional codes in \ac{A-IoT} Rel. 19 \\
\hline
\textbf{Frequency Bands} & Licensed and unlicensed & Regulatory and compatibility constraints & Cellular sub 1 GHz, Wi-Fi, LoRa, mm-wave \\
\hline
\textbf{Standard / Protocol} & EPC Gen2, IEEE 802.11bp AMP, Bluetooth, 3GPP \ac{A-IoT} & Interoperability, deployment feasibility & Cellular and WiFi \ac{ZED} currently being standardized, Bluetooth \ac{A-IoT} gaining traction \\
\hline
\end{tabularx}
\label{table:ambient_iot_parameters}
\end{table*}

\section{\acp{XG-ZED} and  Relation to  \ac{AmBC}}

Figure~\ref{fig:Multi_Static} illustrates the concept of \acp{XG-ZED}, operating in a multi-generation cellular network where radar systems are also present. 
The proposed generation-agnostic \ac{ZED} framework differs fundamentally from \ac{3GPP} \ac{A-IoT}: in \ac{A-IoT}, a continuous RF wave is transmitted to provide power and a modulable signal simultaneously, and dedicated uplink resources are reserved specifically for backscatter transmissions. While this design ensures predictable performance, it requires explicit network support and incurs a continuous energy cost for RF powering. In contrast, \acp{XG-ZED} operate opportunistically and in-band, sharing spectrum and illumination with existing signals \emph{without} needing dedicated resources. 

\acp{XG-ZED} can be classified along several complementary dimensions depending on their role and operating context. A fundamental distinction is between monostatic and bistatic architectures (see again Fig.~\ref{fig:Multi_Static}), where illumination and reception are either co-located or separated. Beyond this, \acp{XG-ZED} can be categorized by their integration into ambient IoT ecosystems, their operative conditions (e.g., energy availability, communication range, and spectral environment), and their realization as chip-aided or chipless tags. Table~\ref{table:ambient_iot_parameters} summarizes the main techniques studied in ambient backscatter communication. These aspects provide the foundation for \acp{XG-ZED}, but must be reconsidered in light of generation-agnostic operation, heterogeneous waveforms, and sustainable large-scale deployment. The following subsections build on this foundation by discussing \ac{XG-ZED} classifications within the ambient IoT landscape, their operative conditions, and their realization as chip-aided or chipless devices.

\subsection{\ac{XG-ZED} Mapping into \ac{A-IoT} }

\ac{3GPP} defines three classes of \ac{A-IoT} devices based on their energy storage and signal generation capabilities. Class A devices have no energy storage and no independent signal generation so that communication relies entirely on backscattering transmission. Class B devices are equipped with limited energy storage but still lack independent RF signal generation, relying mainly on backscatter. Class C devices include both energy storage and independent signal generation, employing active RF components for transmission.

The \acp{XG-ZED} considered in this work correspond most closely to Classes~A and~B: they operate solely via backscatter, with or without a small energy buffer, but without any active RF transmission. In contrast to \ac{A-IoT}, however, we propose a \emph{generation-agnostic} framework based on \acp{AmBC}, whereby devices opportunistically reuse illumination from existing transmissions in the environment, rather than relying on a dedicated \ac{CW} source. Cellular uplink and downlink signals, as well as radar waveforms, already provide structured wideband illumination that can be leveraged simultaneously for communication and localization. Unlike conventional backscatter systems with a dedicated illuminator, \ac{AmBC} shares spectrum and power with the host waveform, thereby achieving high spectral and energy efficiency.


\subsection{Operating Conditions and Receiver Design for \acp{XG-ZED}}

The operating conditions of XG-ZEDs are shaped both by the propagation environment (frequency band and deployment geometry) and by the receiver design.
At \ac{mmWave} frequencies, signals experience severe free-space loss, which is typically compensated by using high-gain antennas that ensure a stable link connection in real operational scenarios. In this case, \acp{XG-ZED} can integrate a retrodirective antenna design, capable of generating a directive beam that follows the direction of arrival of the illuminating signal, maximizing the efficiency of the backscattering link. This favors monostatic deployments, where the transmitter and receiver are co-located. In contrast, sub-6\,GHz bands provide more favorable propagation conditions and support low-gain, omnidirectional antennas, which are well suited for bistatic setups where illumination is provided by either the \ac{UE} or the \ac{BS}. In such bistatic scenarios, the direct path signal is typically orders of magnitude stronger than the one scattered by \ac{XG-ZED}, necessitating mechanisms for direct-path cancellation. 


While \acp{XG-ZED} themselves are illumination-source agnostic, the receiver design is heavily dependent on the signal waveform. Ambient illumination in cellular and radar systems is typically intermittent and structured, making simple envelope-detector receivers inadequate \cite{kellogg2014wi}. Therefore, the receiver must leverage its knowledge of waveform structure for synchronization and coherent detection. This is achieved by using embedded synchronization sequences and pilot/reference signals. Furthermore, advanced receivers may even re-encode decoded data symbols to refine channel estimation. Once the direct channel is reliably estimated, the \ac{XG-ZED}-induced modulation can be detected as a perturbation of the channel response.

\subsection{Chip-Aided and Chipless \acp{XG-ZED}}

Another important classification separates chip-aided from chipless \ac{XG-ZED} designs.
Chip-aided \acp{ZED} incorporate a lightweight integrated circuit that actively controls their backscattering properties, typically by switching between two or more impedance states of the embedded scattering units. This enables digital modulation of the incident signal and provides substantial flexibility: \acp{ZED} can dynamically adapt their modulation schemes to align with different wireless standards, improving robustness and coexistence in heterogeneous spectrum environments. 
Recent circuit architectures, such as \cite{Anxionnat2025}, have shown that chip-aided designs can efficiently exploit LTE reference signals while operating at extremely low power levels, underscoring their strong potential for long-range and energy-efficient \ac{AmBC}. 
Moreover, the integration of lightweight artificial intelligence frameworks such as TinyML opens new opportunities, allowing \acp{ZED} to autonomously adjust modulation strategies and duty cycles to the surrounding spectral conditions and available harvested energy, thereby sustaining resilient performance in dynamic environments.


In contrast, chipless \acp{ZED} rely purely on passive resonant structures (e.g., dipoles, slots, or metamaterial-inspired elements) that imprint a characteristic response on the spectrum of the illuminating signal. In frequency-domain designs, each resonator produces a notch, and the ensemble forms a spectral “barcode” that can be decoded by a wideband receiver, while time-domain approaches use engineered discontinuities in transmission lines to encode information in reflection delays. Beyond communication, chipless \acp{ZED} can also serve as passive sensors by embedding measurand-sensitive materials into resonant structures, whose properties shift in response to environmental changes such as temperature or humidity. Differential architectures or polarimetric backscattering techniques can improve robustness and reading range \cite{Djilani24}, enabling ultra-low-cost, maintenance-free sensing. Potential applications span smart agriculture, structural health monitoring, industrial IoT, and healthcare, where low-cost printed or wearable chipless sensors offer attractive solutions for disposable or long-term monitoring \cite{Ref13_CEA}.

\section{\acp{XG-ZED} Backscattering Communication, Protocols, and Network Architecture}

Building on the classification and design aspects of \acp{XG-ZED} discussed in Section II, we now turn to the communication protocols and network integration aspects. We first review backscatter protocols for ambient \ac{IoT} and \ac{AmBC} systems, before examining how \acp{XG-ZED} can be incorporated into emerging O-RAN and \ac{ISAC} frameworks. Finally, we analyze representative mono- and bistatic link budgets to quantify achievable coverage and performance.

\subsection{Protocols for Ambient Backscatter Communication}

\begin{figure}
\centerline{\includegraphics[width=\columnwidth, trim=0cm 7.5cm 0cm 1cm, clip]{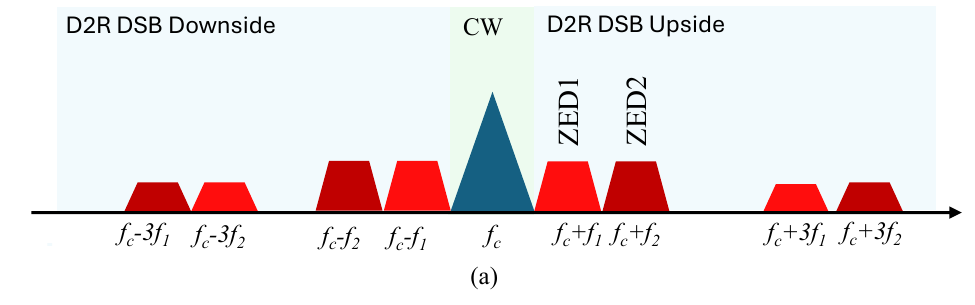}}

\vspace{0.5cm} 

\centerline{\includegraphics[width=\columnwidth, trim=0cm 0cm 0cm 7cm, clip]{Figures/A-IoT_spectrum.pdf}}
    \caption{Spectrum use in ambient IoT and AmBC systems. (a) In conventional A-IoT with a dedicated CW, backscatter devices generate double sidebands (D2R) around the carrier. (b) A-IoT operation with reserved cellular resources, where DL and UL bands are explicitly allocated for R2D and D2R links. (c) AmBC spectrum sharing, where ZEDs reuse existing DL/UL reference signals and their backscattered messages can be simultaneously decoded at both BS and UE.}
    \label{fig:SpectrumRepeated}
\end{figure}

In traditional \ac{A-IoT} systems, generating a dedicated continuous \ac{CW} reduces overall energy efficiency by increasing infrastructure power consumption. An alternative approach based on in-band \acp{AmBC} has been proposed in \cite{jantti2025integration} to eliminate the need for a dedicated \ac{CW}. In this scheme, devices are instead illuminated by existing reference signals, such as \ac{CRS} in the downlink or \ac{SRS} in the uplink.
In the \ac{A-IoT} configuration, resource blocks can be explicitly reserved: one in the downlink for \ac{R2D} communication and another in the uplink for \ac{D2R} communication, as shown in Fig.~\ref{fig:SpectrumRepeated}(b). In contrast, in the \ac{AmBC} case, illustrated in Fig.~\ref{fig:SpectrumRepeated}(c), the \ac{D2R} links share spectrum with both the cellular uplink and downlink, allowing the backscattered message to be simultaneously readable by both the \ac{BS} and the \ac{UE}. In both \ac{A-IoT} and \ac{AmBC} cases, the modulated signal exhibits the spectral characteristics shown in Fig.~\ref{fig:SpectrumRepeated}(a). However, the \ac{AmBC} signal has much smaller frequency shifts, and the spectrum is repeated around each subcarrier. These small shifts result in correspondingly long symbol durations, which allow the receiver channel equalizer to effectively track the multipath components introduced by the backscatter devices. This tracking ensures that interference remains negligible. The primary drawback of this approach is a lower achievable data rate: while several kilobits per second are feasible in the \ac{A-IoT} case, the rate in the \ac{AmBC} case is typically limited to tens or hundreds of bits per second.


\subsection{Integration of \acp{XG-ZED} into O-RAN}
Backscatter messages can be read either at the \ac{UE}, at the \ac{gNB}, or simultaneously at both. One possible realization is to configure \acp{UE} to transmit \ac{SRS}, which are available in LTE and NR and are expected to continue in {6G}. The \ac{gNB} can then use its channel estimator to capture both the direct \ac{UE} response and the perturbations introduced by nearby backscatter devices. With suitable modifications, this function can be extended to demodulate backscatter messages.

\begin{figure}
    \centering
    \includegraphics[width=0.99\linewidth, trim=5.5cm 1cm 6cm 2cm, clip]{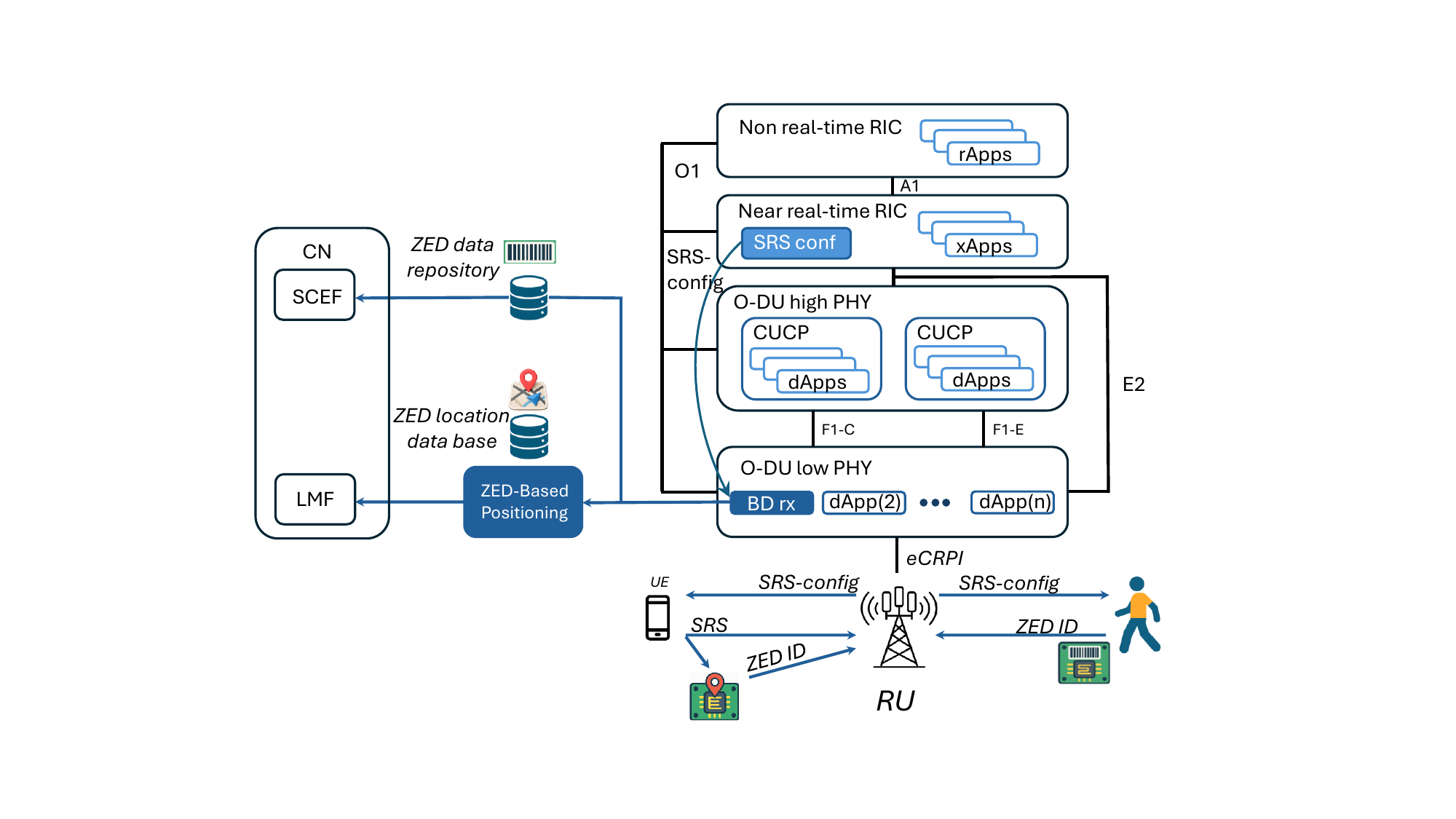}
    \caption{ Integration of XG-ZEDs into the O-RAN architecture. The SMO layer configures UE to transmit SRSs, which illuminate nearby ZEDs. Each ZED modulates the reflected signal to embed a ZED identifier (ZED ID) or payload data. The radio unit (RU) and the distributed unit (O-DU) forward the received signals toward the near-real-time RAN intelligent controller (near-RT RIC), where distributed applications (dApps) decode the embedded information.  The service capability exposure function (SCEF) exposes decoded ZED information to external applications, while the LMF processes ZED-based measurements to support positioning services.  }
    \label{fig:O-RAN-integration}
\end{figure}

Alternatively, the receiver can be implemented within the \ac{O-RAN} framework. The O-RAN Alliance is considering the addition of \acp{dAPP} to the \ac{RIC} specifications. Such \acp{dAPP} would run at the \ac{O-RAN} \ac{DU} (O-DU), where they have access to baseband IQ samples, while transmission of the \ac{SRS} can be scheduled through non-real-time \ac{RIC} functions. Once decoded, the backscatter messages may be delivered to the \ac{CN} through \ac{NAS} signaling, with location information provided to the \ac{LMF} or data payloads routed via existing \ac{A-IoT} functions. Fig.~\ref{fig:O-RAN-integration} illustrates this integration flow within the \ac{O-RAN} architecture, including the roles of the \ac{SMO}, \ac{RIC}, O-DU, and \ac{CN}.

The resulting measurements can be exposed either through dedicated \ac{ISAC} functions (e.g., the \ac{SeCF} and \ac{SPF}) or through existing \ac{CN} functions, demonstrating that \ac{XG-ZED} reception can be naturally aligned with ongoing ISAC standardization efforts. Beyond \ac{O-RAN} integration, practical support for backscatter communication may also be introduced at different levels of the cellular stack: either as incremental extensions of current cellular reference signals and channel estimators, or through future \ac{ISAC}-enabled base stations that unify communication and sensing in a single framework.

\subsection{Link Budget Analysis}

\begin{figure*}[t]
    \centering
    \begin{subfigure}[b]{0.31\textwidth}
        \centering
        \includegraphics[width=\linewidth, trim=6cm 4cm 6cm 3.5cm, clip]{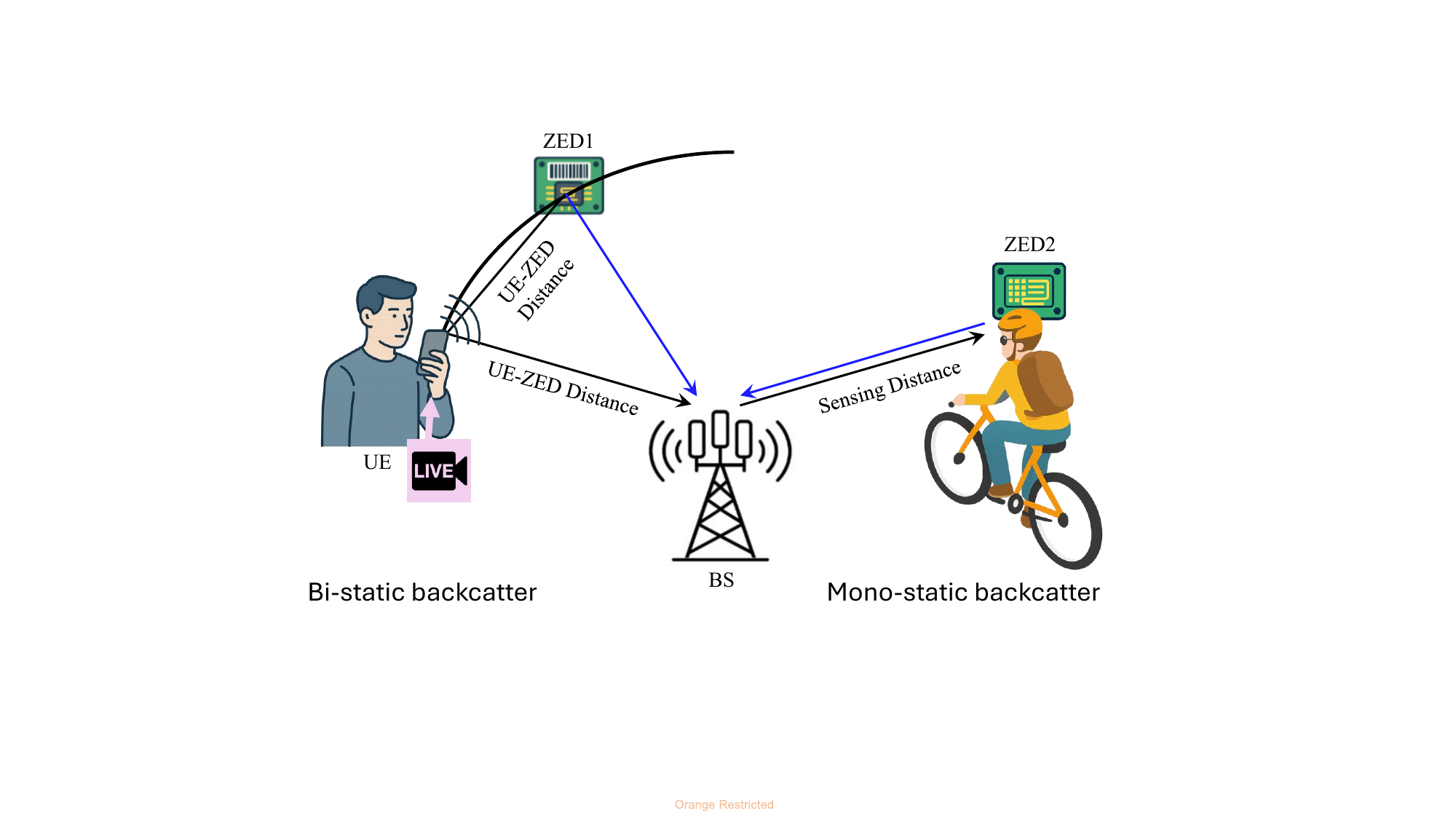}
        \caption{}
        \label{fig:sub_a}
    \end{subfigure}
    \hfill
        \begin{subfigure}[b]{0.43\textwidth}
        \centering
        \includegraphics[width=\linewidth]{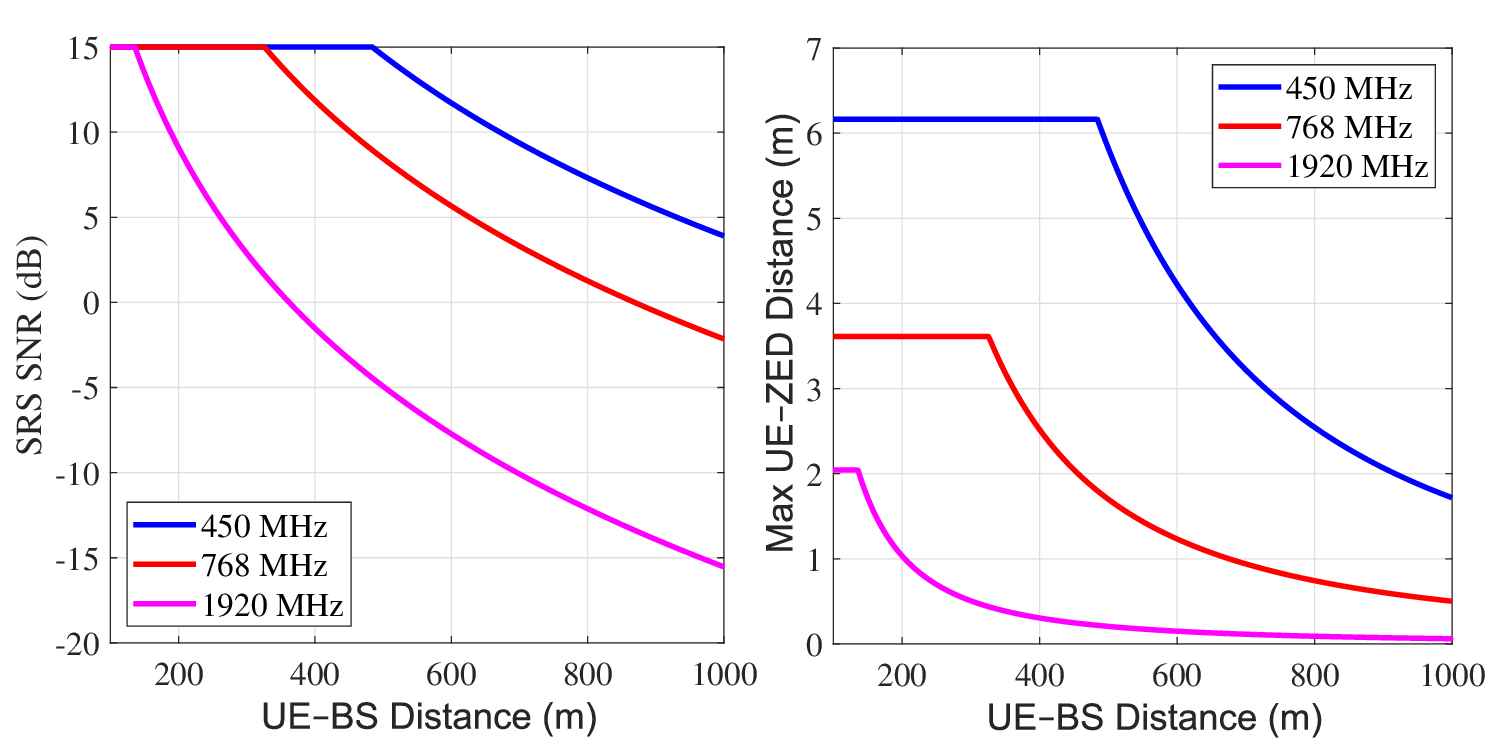}
        \caption{}
        \label{fig:sub_b}
    \end{subfigure}
    \hfill
     \begin{subfigure}[b]{0.247\textwidth}
        \centering
        \includegraphics[width=\linewidth]{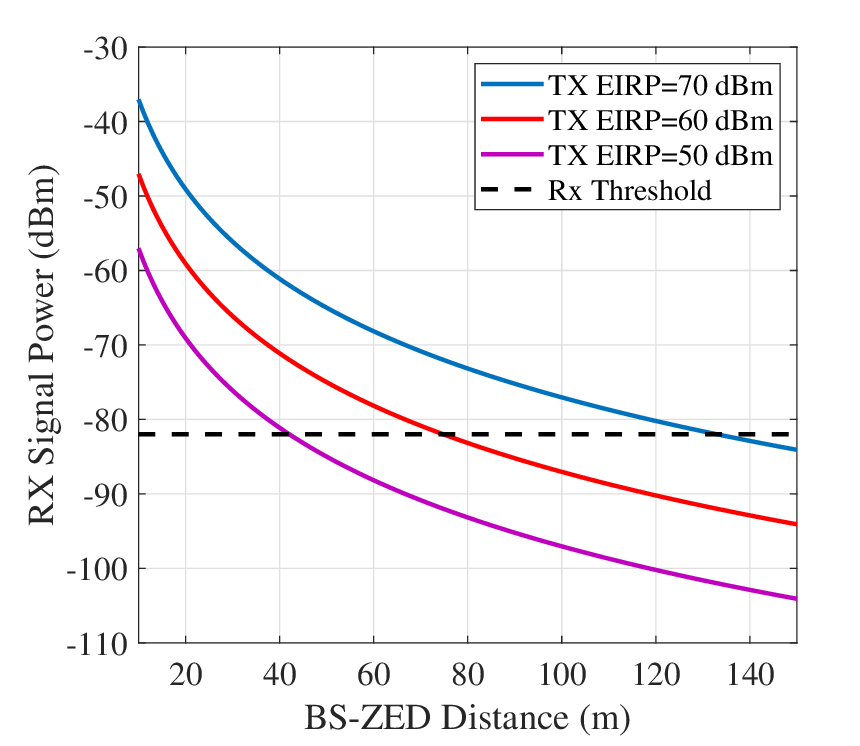}
        \caption{}
        \label{fig:sub_c}
    \end{subfigure}

    \caption{Scenario analysis: (a) bistatic and monostatic backscatter communications, (b) uplink bistatic case: \ac{SNR} for \ac{SRS} (left) and achievable distance between \ac{UE} and \ac{ZED} (right), and (c) link budget for downlink monostatic case.}
    \label{fig:link-budget}
\end{figure*}

To assess the practical feasibility of \ac{XG-ZED} communication, it is essential to analyze the link budget under representative propagation conditions, contrasting sub-6 GHz with mmWave. 
We first consider the bistatic scenario illustrated in Fig.~\ref{fig:link-budget}\subref{fig:sub_a} (left), where the UE and the ZED are equidistant from the BS. Large-scale path loss between the UE and BS, as well as between the ZED and BS, is modeled with the Okumura–Hata suburban path loss model, while the UE–ZED link is assumed line-of-sight and follows free-space loss. The UE transmits \acp{SRS} under BS-controlled power adjustment, targeting an SNR of 15 dB at the BS whenever transmit power (up to 23 dBm) allows. For distant UEs, this target cannot be met, and the received SRS quality degrades, as shown in Fig.~\ref{fig:link-budget}\subref{fig:sub_b}. The ZED introduces an additional 6 dB modulation loss, and reliable backscatter communication requires an SNR of about 1.5 dB, consistent with low-rate coded FSK.
The BS receiver benefits from processing gain, since a single ZED symbol spans many SRS resources across subcarriers, antennas, and OFDM symbols. For LTE SRS configurations, this gain is roughly 31 dB for 10 MHz bandwidth and 34 dB for 20 MHz bandwidth. Representative cases at 450 MHz, 768 MHz, and 1920 MHz show that the maximum ZED reading distance is  dependent on both the carrier frequency and the UE–BS separation, with higher frequencies supporting shorter ranges.

At frequencies above 6 GHz, and especially in the mmWave range, ZEDs can exploit array gain from retrodirective antennas, such as Van-Atta arrays, for monostatic backscattering, Fig.~\ref{fig:link-budget}\subref{fig:sub_a} (right). In the considered 5G NR mmWave downlink scenario, FCC regulations allow up to 75 dBm EIRP per 100 MHz bandwidth. Consider a ZED radar cross-section of –10 dBsm and a 6 dB modulation loss at ZED, together with a 7 dB receiver noise figure and 100 MHz bandwidth. Assuming a 5 dB SNR requirement, the receiver threshold is about –82 dBm. Under these conditions, the results in Fig.~\ref{fig:link-budget}\subref{fig:sub_c} indicate that communication distances well above 100 meters are feasible.

\section{\ac{XG-ZED}-Assisted Sensing, Tracking, and Localization}

In addition to enabling ultra-low-power communication, \acp{XG-ZED} can act as cooperative elements that enhance sensing and localization. By serving as controllable reflectors or identifiable landmarks, \acp{ZED} augment the capabilities of \ac{ISAC} systems, support \ac{R-SLAM}, and provide low-cost anchors for non-line-of-sight positioning. This section outlines these emerging application scenarios and illustrates how \acp{ZED} can complement conventional infrastructure to improve situational awareness.
\subsection{Communication-Centric \ac{ISAC}}

\ac{ISAC} is expected to become a cornerstone of {6G} networks, enabling wireless infrastructures to simultaneously provide data connectivity and radar-like environmental awareness. A compelling vision for \ac{ISAC} lies in the integration of \acp{ZED} as backscatterers in outdoor environments \cite{xie2024power}. Instead of relying on dedicated excitation, these \acp{ZED} could modulate ambient {5G} and future {6G} waveforms, effectively turning everyday objects and vulnerable road users (cyclists, pedestrians, autonomous delivery robots, etc...) into active, distinguishable targets for base stations. When combined with \ac{V2X} technologies, \acp{ZED} could enable real-time detection and classification of road users, dramatically enhancing traffic safety in complex urban scenarios. Since \ac{ISAC} signals are jointly designed for high-resolution sensing and reliable communication, they inherently provide both the reference structures needed for coherent backscatter detection and the required bandwidth. A chip-enabled \ac{ZED} produces a controlled modulation in the sensing signal return, which is analogous to a micro-Doppler signature. Conversely, a chipless \ac{ZED} is identified by its resonant notch-coded response. Thus, sensing systems are capable of both tracking the \ac{ZED} in space and extracting its encoded information. This not only augments the sensing capabilities of ISAC systems but also enables a unified communication-sensing fabric where safety-critical information is embedded directly into the wireless channel.  

\subsection{Radio-SLAM}

The {6G} vision foresees \acp{UE} capable of indoor perception and autonomous navigation in unknown and challenging environments. This entails \acp{UE} autonomously constructing a digital map of their surroundings and performing self-localization using \ac{SLAM} techniques \cite{PaGuGuDeDa2020}. A promising approach to enable this capability is \ac{R-SLAM} (\emph{personal radar}\cite{GuGuDa2016}), which extends \ac{SLAM} into the radio domain. By exploiting high-frequency bands and electronically tunable antennas, \ac{UE}s can scan the environment with narrow beams and process echoes to estimate range, angle, and Doppler shift. 
This supports stand-alone, infrastructure-free mapping and localization, and can complement ISAC-enabled networks to enhance positioning. Importantly, these functions can be realized with standard UE radio hardware. Work in \cite{Munoz25} provides a validated 18–40 GHz polarimetric backscatter and ray-tracing framework for industrial sensing/ISAC.
\begin{figure}
    \centering
    \includegraphics[width=0.99\linewidth]{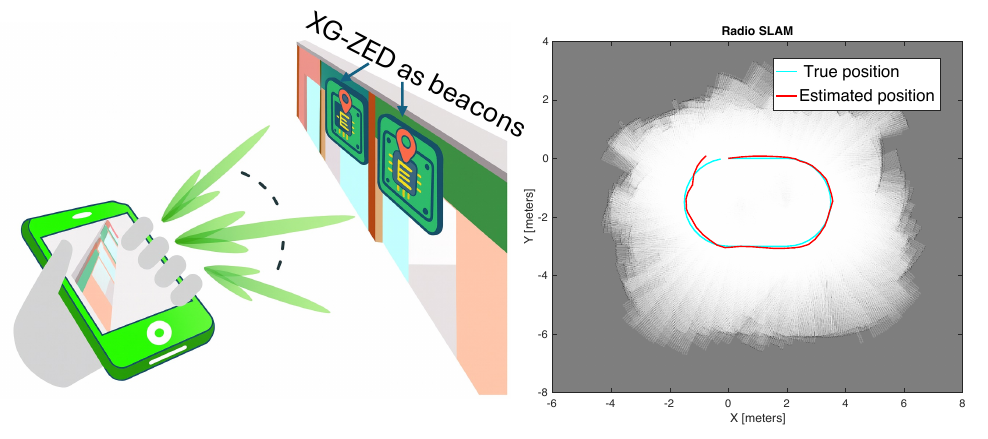}
    \caption{Left:  \acp{XG-ZED} can serve as known landmarks, improving localization and mapping performance. Right: Estimated trajectory and map of a MIMO personal radar moving in an indoor environment equipped.
}
    \label{fig:SLAM-DD}
\end{figure}
However, significant challenges remain: low-reflectivity materials are hard to detect, while highly reflective ones may deflect beams away from the source at oblique incidence angles, causing incomplete maps and reduced accuracy \cite{LoPaGuGuDeDa2023}. To address this, \acp{ZED} can serve as passive  (battery free) radar-visible landmarks (see Fig.~\ref{fig:SLAM-DD}): exploiting retrodirectivity, they can be designed to reflect signals back to the source, acting as artificial and easy-to-identify anchors. Placed strategically (e.g., at corners or doorways), \acp{ZED} serve as battery-free, infrastructure-less reference points that enhance \ac{SLAM} accuracy without adding complexity.

\subsection{Non-line-of-sight Positioning }

Backscatter devices can further serve as reference points for positioning, thereby enabling non-line-of-sight (NLoS) localization in cellular systems. Each \ac{ZED} can transmit a unique identifier, allowing the receiver to associate the detected backscatter signal with the known location of that device. The simplest use case mirrors Bluetooth beaconing, where the presence of a backscatter signal implies proximity, and the \ac{UE} position is inferred from the set of detected \acp{ZED}, as recently demonstrated in real-time using ambient \ac{SRS} signals transmitted by commercial LTE \acp{UE}~\cite{elsanhoury2025zero}. 
More advanced schemes may exploit the relative power of the backscattered path with respect to the direct path as a coarse indicator of distance, or even attempt delay estimation between the direct and backscattered components.  
To be beneficial in practice, the deployment of \acp{ZED} must be relatively dense, since the reliable reading distance between a \ac{UE} and a nearby \ac{ZED} is typically only on the order of a few meters. This makes such devices most attractive in indoor and dense urban environments, where elements of the existing infrastructure (such as walls, ceilings, lamp posts, or street furniture) can host low-cost passive tags that serve as anchors.

\section{Challenges to Realizing \acp{XG-ZED}}

The integration of \acp{XG-ZED} into future wireless systems promises transformative benefits for sustainable connectivity, large-scale sensing, and ubiquitous localization. Yet, realizing this vision requires addressing several open challenges that span technical, environmental, and regulatory domains. These challenges also point toward promising research directions,  outlined below.
\begin{itemize}
   
\item \textbf{Scalability in dense deployments}: Future networks are expected to support millions of XG-\acp{ZED}, often operating in extremely dense scenarios. Ensuring scalability under these conditions is non-trivial, particularly for chipless \acp{ZED}, which face the additional challenge of encoding long identifiers through electromagnetic signatures.

\item \textbf{Spectral coexistence}:  Avoiding harmful interference is essential to guarantee both the reliability of XG-\ac{ZED} operations and the protection of licensed spectrum users. This requires advanced spectrum management strategies and the adoption of coexistence-friendly modulation, coding, and access techniques.

\item \textbf{Localization accuracy versus network complexity}:  Fine-grained positioning often requires multiple observation points or antenna arrays, but both raise cost and practicality issues. An alternative is to use ubiquitous \acp{UE} as cooperative localization agents, though this introduces challenges in synchronization and coordination.

\item \textbf{Lifecycle sustainability}: From an environmental perspective, the sustainability of \acp{ZED} remains only partially assessed. A comprehensive \ac{LCA} is still missing, making it difficult to evaluate their true environmental footprint. Research is therefore needed to promote the use of recyclable materials, extend device lifetime, and ensure that large-scale deployments of \acp{ZED} contribute positively to long-term sustainability goals.

\item \textbf{Standardization and regulatory integration}: For \acp{ZED} to become a widespread and interoperable technology, their integration into existing standardization frameworks is essential. Alignment with bodies such as 3GPP and \ac{O-RAN}, as well as compliance with regulatory policies, will ensure smooth adoption, compatibility with mainstream wireless infrastructures, and safe operation within licensed and unlicensed spectrum.

\item \textbf{Efficient ambient energy harvesting}: Converting RF wireless energy into battery-stored DC power is not immune to the intrinsic losses of the entire harvesting chain.
A possible direction to investigate is based on the concept of opportunistic energy accumulation, based on the principle of virtual RF energy harvesting\cite{Marini2022}.

\item \textbf{Reliable communication with chipless ZEDs}: Fully passive structures encode data via antenna geometry, but their signals are often obscured by environmental clutter. Without a chip, coding capacity is limited, and anti-collision protocols are not feasible. Current \ac{UWB}-based systems using impulse-radar readers face performance bottlenecks, as microstrip resonators lose selectivity at higher frequencies. Future research could explore alternative designs, such as resonant cavities, that could improve spectral efficiency, filtering, and resistance to interference.

 \end{itemize}

\section{Conclusions}
This article introduced the concept of XG-ZEDs, highlighting their potential to enable sustainable, large-scale connectivity by \textit{leveraging ambient cellular signals across generations}. We discussed their classification, operating conditions, and implementation as chip-aided or chipless devices, as well as their integration into cellular protocols, O-RAN frameworks, and ISAC-enabled architectures. Link budget analyses demonstrated the feasibility of both sub-6 GHz and mmWave deployments, while emerging applications in sensing, mapping, and localization underscored their versatility beyond communication. Finally, we outlined key challenges related to scalability, coexistence, sustainability, and standardization that must be addressed to unlock the full potential of XG-ZEDs. By bridging ultra-low-power communication with sensing and positioning, XG-ZEDs represent a promising step toward pervasive, environmentally sustainable IoT in 6G and beyond.
\balance 
\bibliographystyle{IEEEtran}
\bibliography{references}
\vspace{-10mm}
\begin{IEEEbiographynophoto}{Navid Amani} (navid.amani@emickers.com) is the founder of EMickers AB, Sweden, specializing in ultra-low power massive connectivity for wireless communication and sensing with a strong emphasis on sustainability. 
\end{IEEEbiographynophoto}
\vspace{-1.2cm}
\begin{IEEEbiographynophoto}{Filiberto Bilotti} (bilotti@uniroma3.it) is a Full Professor of Engineering Electromagnetics and the Director of the Antennas and Metamaterials Research Laboratory at ROMA TRE University, Italy. His research interests are mainly related to metamaterials and metasurfaces. He is a Fellow of the IEEE.
\end{IEEEbiographynophoto}
\vspace{-1.2cm}
\begin{IEEEbiographynophoto}{Davide Dardari} (davide.dardari@unibo.it) is a Full Professor at the University of Bologna, Italy. 
His interests are in wireless communications, localization techniques, smart radio environments, and distributed signal processing. 
He is a Fellow of the IEEE.
\end{IEEEbiographynophoto}
\vspace{-1.2cm}
\begin{IEEEbiographynophoto}{Raffaele D'Errico} (raffaele.derrico@cea.fr)  is a senior scientist at CEA-LETI, France. His research interests concern propagation channel sounding and modelling and 6G/5G networks.
\end{IEEEbiographynophoto}
\vspace{-1.2cm}
\begin{IEEEbiographynophoto}{Riku Jäntti} (riku.jantti@aalto.fi)   is a Full Professor of Communications Engineering at Aalto University School of Electrical Engineering, Finland. 
\end{IEEEbiographynophoto}
\vspace{-1.2cm}
\begin{IEEEbiographynophoto}{Gianni Pasolini} (gianni.pasolini@unibo.it) is an Associate Professor at the University of Bologna, specializing in wireless communications, digital signal processing, and IoT technologies.
\end{IEEEbiographynophoto} 
\vspace{-1.2cm}
\begin{IEEEbiographynophoto}{Dinh-Thuy Phan-Huy} (dinhthuy.phanhuy@orange.com)
  is
the head of Orange Expertise/ Networks of the Future and a
research project manager on backscattering at Orange/Innovation/Networks, France.
\end{IEEEbiographynophoto}
\vspace{-1.2cm}
\begin{IEEEbiographynophoto}{Davide Ramaccia} (davide.ramaccia@uniroma3.it) is an Assistant Professor of EM field theory at the Department of Industial, Electronic and Mechanical Engineering of RomaTre university. His research interests include metamaterial and metasurfaces.
\end{IEEEbiographynophoto}
\vspace{-1.2cm}
\begin{IEEEbiographynophoto}{Olivier Rance} (olivier.rance@cea.fr) is an RF Engineer with CEA/Electronic for Information Technology Laboratory (LETI), Grenoble.
\end{IEEEbiographynophoto}
\vspace{-1.2cm}
\begin{IEEEbiographynophoto}{Henk Wymeersch} (henkw@chalmers.se) is a Professor at Chalmers University of Technology, focusing on radio localization and sensing.
\end{IEEEbiographynophoto}


\end{document}